\documentstyle[12pt]{article}
\begin{document}

\begin{titlepage}
\title{Sphalerons and large order behaviour of perturbation theory
in lower dimension}

\author{
   V. A. Rubakov and  O. Yu. Shvedov \footnotemark \\
{\small{\em Institute for Nuclear Research of the Russian Academy
of Sciences, }}\\
{\small{\em 60th October Anniversary prospect, 7a, Moscow 117312}}\\
}

\end{titlepage}
\maketitle

\begin{center}
{\bf Abstract}
\end{center}

Sphalerons -- unstable static solutions of classical field equations
in $(d+1)$-dimensional space-time -- may be viewed as euclidean solutions
in $d$ dimensions. We discuss their role in the large order asymptotics
of the perturbation theory. Specifically, we calculate their contribution
to the large order behaviour of the ground state energy in a quantum
mechanical model. When the number of negative modes is odd, single sphaleron
contribution dominates, while this contribution vanishes when the number
of negative modes is even. These results are confirmed by numerical
calculations.

 \footnotetext{ e-mail addresses:rubakov@inucres.msk.su,
 olshv@inucres.msk.su }


\newpage

\section{Introduction}

   Many smooth, finite action classical solutions to euclidean field
equations describe semiclassical tunneling transitions. Ordinary instantons
in Yang-Mills theory \cite{BPST} and their analogs (instantons in
$(1+1)$-dimensional abelian Higgs model, non-linear sigma model, etc. )
 correspond to transitions between degenerate classical vacua
 \cite{H,CDG,JR},
 while bounce solutions \cite{C} are relevant to false vacuum
decay. One of the differences between these two types of solutions is that
the number of negative modes is zero in the former case and one in the
latter.

In this paper we discuss yet another type of euclidean finite action
solutions, namely, sphalerons \cite{M,KM}. Sphalerons are usually
considered as static solutions in $(d+1)$-dimensional field theories where
they determine the height of a barrier between different vacua. However,
they may  be viewed also as (time-dependent) solutions to euclidean field
equations in $d$-dimensional space-time. In  this context we will call these
solutions ''sphinstantons''. We are interested in their role in
 $d$-dimensional field theories.

Examples of sphinstantons are:

i) original sphaleron \cite{M,KM} viewed as the euclidean solution in
3-dimensional Yang-Mills-Higgs theory;

ii) sphaleron in 2-dimensional non-linear sigma model with explicit
symmetry breaking \cite{MW} interpreted as the  euclidean solution in
quantum mechanics of a particle moving on a two-dimensional sphere in
homogeneous gravitational field;

iii) the solution $I^{*}$ in 4-dimensional euclidean Yang-Mills-Higgs
theory, whose existence has been recently advocated in
refs. \cite{K1,K2}; etc.

Like bounces, sphinstantons have negative modes (sphalerons in
 $(d+1)$ dimensions are unstable). At first sight, this implies, that these
negative modes signalize the vacuum instability,
and the sphinstantons
describe the semiclassical vacuum decay.
 However, in known examples there
are no classical field configurations with energy smaller than the energy
of the trivial vacuum. Thus, the trivial vacuum is stable (at least
semiclassically), so the vacuum decay interpretation of sphinstantons is
not possible. The formal reason for this is the absence of a non-trivial
turning point (the slice $t$=const at which time derivatives of the fields
is zero at all ${\bf x}$ ), as opposed to the bounce solutions.

Besides describing tunneling, instantons play another role in field theory
and quantum mechanics. Namely, bounces and
 instanton -- anti-instanton pairs contribute
to the high orders of perturbation theory
\cite{BBW,Lpt,BF,BPZJ}
 (for a review see \cite{ZJ} ).
The contribution of the instanton -- anti-instanton pair to the $k$-th
order of perturbation theory is proportional to
\begin{equation} k!/(2S_{I})^{k},  \label{4*}\end{equation}
where $S_{I}$ is the action of the instanton.
The fact that instanton -- anti-instanton pairs,
rather than single instantons, contribute to
eq. (\ref{4*}) is related to the absence of negative modes around the
instantons. One may expect that sphinstantons also contribute to
high orders of perturbation theory, and their contribution behaves as
\begin{equation} k!/S_{SI}^{k} , \label{4+}\end{equation}
if the number of negative modes is odd, and
\begin{equation} k!/(2S_{SI})^{k}  ,\label{4**}  \end{equation}
if the number of negative modes is even (in the latter case the
contribution should again come from sphinstanton pairs).
Here $S_{SI}$ is the sphinstanton action (equal to the static energy
of the sphaleron in $(d+1)$ dimensions).

The purpose of this paper is to check these expectations, eqs. (\ref{4+})
and (\ref{4**}),  in a simple quantum mechanical model of a particle
moving on a sphere in external potential. As pointed out above, sphinstanton
in this model is the sphaleron in 2-dimensional non-linear sigma model
with explicit symmetry breaking term. We describe the model and its
sphinstanton in sect. 2. In sect. 3 we present detailed calculations of the
high orders of perturbation theory for the ground state energy, making use
of the technique analogous to ref. \cite{ZJ}. In fact, we consider
in sect. 3
a particle on $n$-dimensional sphere, where $n$ need not be integer.We will
see that at all $n$ except for $n=1,3,5,...$, the asymptotics is given by
eq. (\ref{4+}), while at integer odd $n$ (when the sphinstanton has even
number of negative modes), the one-sphinstanton contribution vanishes,
so that the asymptotics is given by eq. (\ref{4**}). In sect. 4 we present
the results of numerical calculations of the high orders of perturbation
theory which agree with the analytical results of sect. 3. Sect. 5 contains
concluding remarks.

\section{The model and sphinstanton}
We consider quantum mechanics of a particle with unit mass moving on
$n$-dimensional sphere in external potential. The classical euclidean action
is
\[ S=\int dt[(d{\bf N}/dt)^{2}+V(N_{1})] ,    \]
where ${\bf N}=(N_{1},...,N_{n+1}),{\bf N}^{2}=1$. We consider the potentials
depending
only on one coordinate $N_{1}$, and assume that it has one minimum at
$N_{1}=-1$, where $V(-1)=0,V^{'}(-1)=1$.
 Then the classical ground state is the point
\[ {\bf N}_{0}=(-1,0,0,...,0) \]
and its classical energy is zero. Notice that both in classical and quantum
mechanics the ground state is unique, and, therefore, stable.

Sphinstanton in this model (cf. ref. \cite{MW})
up to $O(n)$ rotations of
coordinates $N_{2},...,N_{n+1}$ is
\[   N_{1}=\cos \theta (t), N_{2}=\sin \theta (t),\]
where $\theta(t)$ is determined by the following relation
\[  t=\int_{0}^{\theta} \frac{d\theta}{\sqrt{2V(\cos \theta)}}.  \]
As $t$ runs from $-\infty$ to $+\infty$, the particle makes a loop around
the sphere from the south pole through the north pole back to the south
pole (see fig. 1). The action for the sphinstanton is
\[  S_{SI}=\int_{-\pi}^{\pi} \sqrt{2V(\cos \theta)} d\theta  \]
It is obvious that there exist $(n-1)$ negative modes around the sphinstanton
 (cf. ref. \cite{MW}), which correspond to the shifts of the loop in $(n-1)$
directions, see fig. 1.

The evaluation of the contributions of classical solutions to high orders
of perturbation theory is by now standard (see ref. \cite{ZJ}). In this paper
we consider perturbation theory for the ground state energy, the perturbation
parameter being $\hbar$,
\begin{equation}  E_{0}=\hbar E^{(0)}+\hbar^{2}E^{(1)}+...+\hbar^{k+1}
E^{(k)}+...    \label{9*}   \end{equation}
Roughly speaking, $E^{(k)}$ is given by the integral
\[ E^{(k)} \sim  \int \frac{d\hbar D{\bf N}}{2 \pi i\hbar^{k+1}}
\exp(-S/\hbar) ,\]
where the integrals are to be taken by the saddle-point method. The
contribution of the sphinstanton to $E^{(k)}$ is then
\[  E^{(k)} \sim  \int \frac{d\hbar}{2\pi i\hbar^{k+1}}D^{-1/2}
\exp(-S_{SI}/\hbar) , \]
where $D$ is the determinant of fluctuations around the sphinstanton.
To make real contribution to $E^{(k)}$, the sphinstanton should have odd
number of negative modes, so that $D^{1/2}$ is pure imaginary. Then the
asymptotic behaviour of $E^{(k)}$ is given by eq. (\ref{4+}). When the
number of negative modes is even, the contribution of a single sphinstanton
should vanish (otherwise it would be pure imaginary, in contradiction to
the general property of reality of $E^{(k)}$). So, we expect that the
behaviour of $E^{(k)}$ at large $k$ is
\[ E^{(k)} \propto  k!/S_{SI}^{k}  ,\;\;\;\; \mbox{     even } n, \]
\[ E^{(k)} \propto  k!/(2S_{SI})^{k} ,\;\;\;\; \mbox{     odd } n.\]

We will check this property by numerical calculations in sect. 4, and now
we turn to actual calculation of the sphinstanton contribution to
$E^{(k)}$ at arbitrary (not necessarily integer) dimensionality of the
sphere.

\section{Sphinstanton contribution to high orders of perturbation theory}

In this section we evaluate the sphinstanton contribution to the
perturbative expansion of the ground state energy.
Since the potential depends only
on $\theta$, the ground state has zero angular momentum with respect to
other $(n-1)$ angular variables, i.e., the wave function of the ground state
depends only on $\theta$. Thus, we consider the s-wave Hamiltonian
\begin{equation} H=-\frac{\hbar^{2}}{2}\frac{d^{2}}{d\theta^{2}}-
\frac{\hbar^{2}}{2}(n-1)
 \frac{\cos\theta}{\sin\theta}
 \frac{d}{d\theta} + V(\cos \theta) \label{12*}\end{equation}
 The wave function of the ground state should be non-singular at the
south and north poles, so we require
\begin{equation}  \frac{d\psi}{d\theta}(\theta=0)=0
\label{12**}\end{equation}
\begin{equation}  \frac{d\psi}{d\theta}(\theta=\pi)=0
\label{12+}\end{equation}
The Hamiltonian (\ref{12*}) and boundary conditions (\ref{12**}),
 (\ref{12+}) make sense at arbitrary, not necessarily integer $n$. So, we may
generalize the problem by defining the ground state of a particle on
fractionally dimensional sphere as the lowest energy state of the
Hamiltonian (\ref{12*}).

We are interested in $k$-th order of the perturbation theory for the ground
state energy, $E^{(k)}$, see eq. (\ref{9*}). To evaluate this term, it is
convenient to consider the statistical sum
\[  Tr \exp(-\beta H/\hbar)=\int_{\theta(0)=\theta(\beta)} D\theta
\exp(-S/\hbar) \]
The $k$-th order term in the expansion of the statistical sum is formally
written as
\begin{equation} \left( Tr \exp(-\beta H/\hbar)\right)^{(k)}
= \int \frac{d\hbar D\theta}
{2\pi i \hbar^{k+1}} \exp(-S/\hbar),  \label{13*}\end{equation}
where the contour of integration over $\hbar$ runs around the origin
counterclockwise.

Consider now the sphinstanton contribution to the integral
\begin{equation}  (Tr \exp(-\beta H/\hbar))_{SI} =
\int_{\theta(0)=\theta(\beta)} D\theta \exp(-S/\hbar)
\label{13+}\end{equation}
We will see that this contribution behaves as follows,
\begin{equation}   (Tr \exp(-\beta H/\hbar))_{SI}=
\beta \exp(-\beta n/2)iB\hbar^{-\alpha}\exp(-S_{SI}/\hbar) \label{14+}
\end{equation}
where $\alpha$ and $B$ are some functions of $n$. From eq. (\ref{13*})
 it follows that the sphinstanton gives rise to the following
 asymptotic behaviour  at large $k$
\begin{equation} (Tr \exp(-\beta H/\hbar))^{(k)}=
\beta\exp(-\beta n/2)\frac{k!k^{\alpha-1}B}{2\pi S_{SI}^{k+\alpha}}
\label{14**} \end{equation}
On the other hand, at large $\beta$ the statistical sum is saturated by
the ground state, so that
\begin{equation} Tr \exp(-\beta H/\hbar)=
\exp(-\beta(E^{(0)}+\hbar E^{(1)}+...+\hbar^{k}E^{(k)}+...))
\label{14*}\end{equation}
where $E^{(0)}=n/2$. One then expands eq. (\ref{14*}) in $\hbar$ and
makes use of the factorial growth, eq. (\ref{14**}), to obtain
 (cf. ref. \cite{ZJ})
\begin{equation}
E^{(k)}=-\frac{k!k^{\alpha-1}B}{2\pi S_{SI}^{k+\alpha}} \label{14++}
\end{equation}
So, our main problem is to evaluate the sphinstanton contribution to the
functional integral (\ref{13+}).

Before the actual calculation of the sphinstanton contribution, we make a
few comments concerning the integral (\ref{13+}). First,
we wish to make use
of the saddle point technique. The Hamiltonian (\ref{12*}) has, in fact,
a term of order $\hbar$, so it is convenient to write the euclidean action in
the
following form, up to terms of order $\hbar^{2}$,
\[ S=S_{0}+\hbar S_{1} ,\]
where
\[ S_{0}=\int dt (-ip_{\theta}\dot{\theta}+H_{0}(p_{\theta},\theta)) , \]
\[ S_{1}=\int dt H_{1}(p_{\theta},\theta), \]
where
\[ H_{0}=p_{\theta}^{2}/2+V(\cos\theta),  \]
\[ H_{1}=-i\frac{n-1}{2}\frac{\cos\theta}{\sin\theta} p_{\theta} , \]
where $p_{\theta}$ is the momentum  conjugate to $\theta$.
We will consider saddle points of the action $S_{0}$ and treat
$\exp(-S_{1})$ as the pre-exponential factor. Second, the points $\theta=0$
and $\theta=\pi$ are singular points where the semiclassical approximation
does not work. When treating the regions close to these points, we will
have, in particular, to take into account the boundary conditions
 (\ref{12**}), (\ref{12+}). Finally, we first study finite values of $\beta$.
 The relevant classical euclidean trajectory is then periodic in time
 with the period $\beta$. We take the variable $\theta$ to belong to the
 interval $\theta \in (0,\pi)$, so the function $\theta(t)$ has
 discontinuous  time derivatives at $\theta=0$ and $\theta=\pi$ (see fig. 2).
 This will not make a serious problem because we will anyway treat the
 points $\theta=0$ and $\theta=\pi$ ''exactly''.

Let us consider the vicinities of the points $\theta=\pi$ and $\theta=0$
in more detail. We choose some points $\theta_{r}=\pi-r$, $\theta_{s}=s$,
that separate semiclassical and non-semiclassical regions, and take $r$ and
$s$ small enough. Every trajectory close to the classical solution has the
form shown in fig. 2. We may choose the origin of time in such a way that
the trajectory passes the point $\theta_{r}$, running towards $\theta=\pi$,
at $t=0$. Then it passes the point $\theta_{r}$ again at some time
$t=\tau_{1}$, and then passes the point $\theta_{s}$ twice at $t=
\tau_{1}+\tau_{2}$ and $t=\tau_{1}+\tau_{2}+\tau_{3}$, as shown in fig. 2.
Notice that at large $\beta$, the point spends almost all euclidean ''time''
near $\theta=\pi$, so that $\tau_{1}$ is close to $\beta$, while $\tau_{2}$,
$\tau_{3}$ and $(\beta-\tau_{1}-\tau_{2}-\tau_{3})$ remain finite in the
limit $\beta \rightarrow \infty$. Guided by this observation,
we insert the identity
\[
\beta\int d\tau_{1} d\tau_{2} d\tau_{3} \delta(\theta(0)-\theta_{r})
\delta(\theta(\tau_{1})-\theta_{r}) \delta(\theta(\tau_{1}+\tau_{2})-
\theta_{s}) \delta(\theta(\tau_{1}+\tau_{2}+\tau_{3})-\theta_{s})  \]
\[ \times
\dot{\theta}(0)\dot{\theta}(\tau_{1})\dot{\theta}(\tau_{1}+\tau_{2})
\dot{\theta}(\tau_{1}+\tau_{2}+\tau_{3}) =1 \]
 into the integral (\ref{13+}). This identity is valid when integrated with
 the weight invariant under time translations, and the factor $\beta$
 accounts for the time translational invariance. The resulting integral
 is then expressed through Green's function
 \begin{equation}
  G(\theta_{1},\theta_{2};t_{2}-t_{1})=\int_{\theta(t_{1})=\theta_{1},
 \theta(t_{2})=\theta_{2}}\exp(-S/\hbar)D\theta \label{18*}\end{equation}
 Namely, we obtain
 \[
 (Tr \exp(-\beta H/\hbar))_{SI}=\beta \int d\tau_{1} d\tau_{2} d\tau_{3}
 \dot{\theta}(0) \dot{\theta}(\tau_{1}) \dot{\theta}(\tau_{1}+\tau_{2})
 \dot{\theta}(\tau_{1}+\tau_{2}+\tau_{3}) \]
 \begin{equation}    \times
  G(\theta_{r},\theta_{r};\tau_{1}) G(\theta_{r},\theta_{s};\tau_{2})
  G(\theta_{s},\theta_{s};\tau_{3}) G(\theta_{s},\theta_{r};
  \beta-\tau_{1}-\tau_{2}-\tau_{3})     \label{18+}
  \end{equation}

 The integral (\ref{18*}) for Green's functions $G(\theta_{r},\theta_{s};
 \tau_{2})$ and $G(\theta_{s},\theta_{r};\beta-\tau_{1}-\tau_{2}-\tau_{3})$
 should be calculated by the saddle-point (semiclassical) technique. The
 semiclassical expression for Green's function $G(\theta_{r},\theta_{s};
 \tau_{2})$ has the following form \cite{Msl1,Msl2}
 \[
 G(\theta_{r},\theta_{s};\tau_{2})=
 \frac{\exp(-
 \int_{\tau_{1}}^{\tau_{1}+\tau_{2}} H_{1} d\tau)}{\sqrt{2\pi \hbar}}
  \left(\frac{\partial\theta(
 \tau_{1}+\tau_{2},\theta_{r},\dot{\theta}(\tau_{1}))}{\partial\dot{\theta}
 (\tau_{1})}\right)^{-1/2} \]
 \begin{equation}
  \times \exp(-S(\theta_{r},\theta_{s};\tau_{2})/\hbar) \label{18**}
 \end{equation}
 where $\theta(\tau,\theta_{r},\dot{\theta}(\tau_{1}))$ is the classical
 trajectory which starts at $\theta_{r}$ at $\tau=\tau_{1}$ with the
 velocity $\dot{\theta}(\tau_{1})$;
  after taking the derivative in eq. (\ref{18**}) one should choose
 $\dot{\theta}(\tau_{1})$ in such a way that the trajectory passes
 $\theta=\theta_{s}$ at $t=\tau_{1}+\tau_{2}$;
    $S(\theta_{r},\theta_{s};\tau_{2})$
 is the classical action along the latter trajectory. A completely
 analogous expression may be written for
 $G(\theta_{s},\theta_{r};\beta-\tau_{1}- \tau_{2}-\tau_{3})$.

 On the other hand, Green's functions $G(\theta_{r},\theta_{r};\tau_{1})$
 and $G(\theta_{s},\theta_{s};\tau_{3})$ cannot be calculated by the
 saddle point technique: the regions $\theta>\theta_{r}$ and $\theta<\theta_
 {s}$ are not semiclassical. Consider first $G(\theta_{r},\theta_{r};
 \tau_{1})$. Since $\tau_{1} \approx \beta$ at large $\beta$,
 we may approximate
 this Green's function by the ground state contribution to its expansion
 through the eigenfunctions of the Hamiltonian,
 \[ G(\theta_{r},\theta_{r};\tau_{1})=r^{n-1}\psi_{0}(r)\psi_{0}^{*}(r)
 exp(-\tau_{1}E_{0})\]
Furthermore, since $r$ is small, the Hamiltonian at $\pi>\theta>\theta_{r}=
\pi-r$ coincides with its quadratic part,
\[ H_{0}=-\frac{h^{2}}{2}\frac{d^{2}}{dr^{2}}-\frac{h^{2}}{2}
\frac{n-1}{r}\frac{d}{dr}+\frac{r^{2}}{2}  \]
so that the ground state wave function is
\[ \psi_{0}(r)=
\left(\frac{2}{h^{n/2}\Gamma(n/2)}\right)^{1/2}\exp(-r^{2}/2\hbar) \]
Note that this wave function obeys the boundary condition (\ref{12+})
automatically. We obtain finally
\[ G(\theta_{r},\theta_{r};\tau_{1})=\frac{2r^{n-1}\exp(-r^{2}/\hbar)}
{\hbar^{n/2}\Gamma(n/2)}\exp(-\tau_{1}n/2).  \]

It is now straightforward to integrate over $\tau_{1}$ and $\tau_{2}$ in
eq. (\ref{18+}). This integral is again of the saddle-point nature, with the
relevant exponent being
\[ \exp\left(-\frac{S(\theta_{r},\theta_{s};\tau_{2}) +
S(\theta_{s},\theta_{r};\beta-\tau_{1}-\tau_{2}-\tau_{3})}{\hbar}\right) \]
In the limit $\beta \rightarrow \infty$,
the saddle point in $\tau_{1}$ and $\tau_{2}$ is obtained when the two
trajectories connecting $\theta_{r}$ and $\theta_{s}$ have zero energy, i.e.,
when the classical trajectory is the sphinstanton. The pre-exponential
factors are obtained by making use of the following relations:

i) the duration of the trajectory connecting $\theta_{r}$ and $\theta_{s}$
at energy $E$ is
\[ \tau_{2}(E)=- \int_{\theta_{r}}^{\theta_{s}} \frac{d\theta}
{\sqrt{2(E+V(\cos \theta))}}   \]

ii) the expression entering eq. (\ref{18**}) is
\[ \frac{\partial\theta(\tau_{1}+\tau_{2},\theta_{r};\dot{\theta}(\tau_{1}))}
{\partial\dot{\theta}(\tau_{1})}=-\dot{\theta}(\tau_{1}+\tau_{2})\dot{\theta}
(\tau_{1})\left(\frac{d\tau_{2}(E)}{dE}\right)_{E=0} \]

iii)the second derivative of the action with respect to $\tau_{2}$ is
\[ \frac{\partial^{2}S(\theta_{r},\theta_{s};\tau_{2})}{\partial\tau_{2}^{2}}=
-\left(\frac{1}{d\tau_{2}(E)/dE}\right)_{E=0} \]

iv)the integrals $\int H_{1} d\tau$ along the two parts of the sphinstanton
trajectory cancel each other, because $H_{1}$ is linear in momentum
$p_{\theta}$;

v)by the explicit calculation of  $\theta(\tau)$ for the sphinstanton
one finds at small $\theta_{s}$ and small $r=\pi-\theta_{r}$,
\[ r\exp(-\tau_{1}/2)=\pi \exp(-\beta/2) \exp \int_{0}^{\pi} d\theta\left(
\frac{1}{\sqrt{2V(\cos \theta)}} - \frac{1}{\pi-\theta}\right)     \]

Collecting all pre-exponentional factors we write the following expression
for the sphinstanton contribution (\ref{18+}) after the integration over
$\tau_{1}$ and $\tau_{2}$,
\[ (Tr \exp(-\beta H/\hbar))_{SI}=\frac{2\beta \exp(-\beta n/2)}{\hbar^{n/2}
\Gamma(n/2)} \left( \pi \exp\left[\int_{0}^{\pi} d\theta
\left(\frac{1}{\sqrt{2V(\cos\theta)}}
- \frac{1}{\pi -\theta}\right)\right]\right)^{n} \]
\begin{equation} \times \int d\tau_{3} \sqrt{2V(\cos \theta_{s})}
G(\theta_{s},\theta_{s};\tau_{3})  \label{21*} \end{equation}

Notice that the above technique can be used for obtaining high order
behaviour
 of the perturbation theory in simpler models. We show in Appendix that
the known results \cite{BBW,ZJ} for $O(n)$-symmetric single well potentials
are indeed reproduced in this way.

The remaining part of the calculation is the evaluation of the ''sphinstanton
contribution'' to Green's function $G(\theta_{s},\theta_{s};\tau_{3})$. At
$0<\theta<\theta_{s} \ll  1$, the Hamiltonian is approximated as follows
\[H=-\frac{\hbar^{2}}{2}\frac{d^{2}}{d\theta^{2}}-\frac{\hbar^{2}}{2}
\frac{n-1}{\theta}\frac{d}{d\theta}+E_{NP},  \]
where $E_{NP}=V(\cos \theta=1)$ is the potential energy at the north pole.
The exact Green's function of this operator, which obeys
the boundary condition (\ref{12**}), is
\[ G(\theta_{2},\theta_{1};\tau)=\frac{2 \pi^{(n-1)/2} \theta_{2}^{n-1}}
{\Gamma((n-1)/2)(2 \pi \hbar \tau)^{n/2}}\exp\left(-\frac{E_{NP}\tau}{\hbar}-
\frac{\theta_{1}^{2}+\theta_{2}^{2}}{2\hbar \tau}\right) \]
\[ \times\int_{-1}^{1} d\lambda
\exp\left(\frac{\theta_{1}\theta_{2}\lambda}{h\tau}\right)
(1-\lambda^{2})^{(n-3)/2} \]

The extrema of the exponent in the integral over $\lambda$ are end points,
$\lambda=\pm 1$. We keep
the contribution of $\lambda=-1$ only. This contribution corresponds to
trajectories that are reflected by the point $\theta=0$: indeed, the
$\theta$-dependent term in the exponent at $\lambda=-1$ is
\[ (\theta_{1}+\theta_{2})^{2}/(2\hbar\tau)\]
which is preceisely the action for such trajectories. To evaluate the
contribution of $\lambda=-1$, we change the integration contour as shown in
fig. 3 and obtain the ''sphinstanton contribution''
\[ (G(\theta_{s},\theta_{s};\tau_{3}))_{SI}=
-\frac{2i\sin(\pi (n-3)/2))}{\sqrt{2\pi \hbar \tau_{3}}}
\exp\left(-\frac{E_{NP}\tau_{3}}{\hbar}-\frac{2\theta_{s}^{2}}
{\hbar\tau_{3}}\right) \]
We need the integral of this expression over $\tau_{3}$ (see eq. (\ref{21*})).
The integration is straightforwardly performed by the saddle point
technique; we find
\[ \int d\tau_{3} \sqrt{2E_{NP}} (G(\theta_{s},\theta_{s};\tau_{3}))_{SI}=
-2i\cos(\pi n/2) \]
Inserting this expression into eq. (\ref{21*}), we obtain the sphinstanton
contribution to the integral (\ref{13*}) in the form (\ref{14+}) where
\[ \alpha=n/2, \]
\[ B=-\frac{4\pi^{n}}{\Gamma(n/2)}\cos(\pi n/2)
\exp\left(n\int_{0}^{\pi}d\theta \left(\frac{1}{\sqrt{2V(\cos\theta)}}
-\frac{1}{\pi-\theta}\right)\right) \]
According to eq. (\ref{14++}), the high order behaviour of the ground state
energy is finally
 \begin{equation} E^{(k)}=\frac{k!k^{n/2-1}}{S_{SI}^{k+n/2}}\cos(\pi n/2)
\frac{2\pi^{n-1}}{\Gamma(n/2)}
\exp\left(n\int_{0}^{\pi}d\theta \left(\frac{1}{
\sqrt{2V(\cos\theta)}}-\frac{1}{\pi-\theta}\right)\right)
  \label{24*}\end{equation}
An interesting feature of this formula is that it is zero at integer odd
$n$, i.e., the sphinstanton does not contribute to the high order behaviour
of the perturbation theory at these $n$. This confirms our expectations
 (sect. 2) based on  counting of the negative modes.

\section{Numerical results}
 In this section we confirm the results of sect. 3 by numerical  calculations
of the perturbative expansion of the ground state energy. The numerical study
is based on the following method \cite{BW}. The wave function of the ground
state is expanded in $\hbar$,
\[ \psi=\sum \psi^{(k)}\hbar^{k} \]
where
\[ \psi^{(k)}=\sum_{l=0}^{2k} A_{k,l}x^{2l}\exp(-x^{2}/2), \]
\[ x=(\pi-\theta)/\sqrt{\hbar} \]
The coefficients $A_{k,l}$ obey recursive relations which enable one to
develop the numerical procedure.

We perform numerical calculations for the particular form of the  potential,
\begin{equation}
V(\cos\theta)=\cos\theta+1   \label{25*}
\end{equation}
This choice corresponds to a particle moving on a sphere in homogeneous
gravitational field. The recursive relations in this case have the
following form
\[ -(l+1)(2l+n)A_{k,l+1}+\sum_{m=2}^{l}(-1)^{m+1}A_{k-m+1,l-m}/m!
 -\sum_{m=1}^{k} E^{(m)}A_{k-m,l} \]
\begin{equation}
+2lA_{k,l}+\sum_{m=1}^{l}f_{m}A_{k-m,l-m}-
2\sum_{m=1}^{l}f_{m}(l-m+1)A_{k-m,l-m+1}=0 \label{26*}
\end{equation}
where
\[f_{m}=-\frac{(n-1)2^{2m}|B_{2m}|}{2(2m)!} \]
are the coefficients of the expansion of
\[ \frac{(n-1)x\cos x}{2\sin x} = \sum f_{k}x^{2k} \]
($B_{2m}$ are Bernoulli numbers), and by definition
\[ A_{0,0}=1, \]
\[ A_{k,0}=0\;,\;k>0 \]
\[ A_{k,l}=0\;,\;l<0 \;,\; l>2k \]
As before, $E^{(k)}$ is the $k$-th coefficient in the expansion of the ground
state energy.

Eq. (\ref{26*}) defines the numerical procedure
for evaluation of both
$A_{k,l}$ and $E^{(k)}$.
We are interested in the latter quantity, which
should  be compared to eq. (\ref{24*}). For  the particular choice of the
potential, eq. (\ref{25*}), we have
\[ S_{SI}=8 \]
and
\[ \int_{0}^{\pi} d\theta \left(\frac{1}{\sqrt{2V(\cos\theta)}}-
\frac{1}{\pi- \theta}\right)=\ln (4/\pi)  \]
So, we find from  eq. (\ref{25*}) that the following quantity,
\begin{equation}
D_{k}=\frac{8^{k}E^{(k)}}{k!k^{n/2-1}} \label{14A}
\end{equation}
tends to constant as $k \rightarrow \infty$,
\begin{equation}
 \lim_{k \rightarrow \infty} D_{k}=D=\frac{2^{n/2+1}\cos(\pi n/2)}{\pi
\Gamma(n/2)} \label{27*} \end{equation}

The results of our calculations are shown in figs. 4-6.
In fig. 4a,b,c we plot $D_{k}$ at various $n$. It is clear that they indeed
tend to  the values given by eq. (\ref{27*}) which are shown by dashed
lines.

In fig. 5 we plot $D_{k}$ versus $k$ (in logarithmic scale) at integer odd
$n$. We see from fig. 5 that sphinstantons indeed do not contribute to the
high orders of perturbation theory at these  $n$ ($D_{k}$  exponentially
tend to zero at large $k$).

Fig. 6 shows that the  behaviour of $E^{(k)}$ at large $k$, and $n$ close
to odd number is consistent with the sum of one sphinstanton and
sphinstanton -- anti-sphinstanton contributions,
\[ E^{(k)}=C_{1}k^{\alpha_{1}}\cos(\pi n/2)k!/(S_{SI}^{k})+
C_{2}k^{\alpha_{2}}k!/(2S_{SI})^{k}  \]
where $C_{1,2}(n)$ and $\alpha_{1,2}(n)$ are regular at $n=1,3,5,...$.
Namely, at relatively small $k$  the ratio
\begin{equation}
F_{k}=E^{(k+1)}/(kE^{(k)}) \label{15B}
\end{equation}
is close to $1/(2S_{SI})=1/16$
 (the sphinstanton pair dominates) but at
\[ k>>|\ln(\cos(\pi n/2))|\] the single sphinstanton
 wins, and $F_{k}$ becomes
equal to $1/S_{SI}=1/8$.

Thus, our numerical  results show that it is indeed the single
sphinstanton that determines the high order behaviour of the perturbation
theory at $n \neq 1,3,5,... $, while at integer odd $n$ the behaviour is
quite different and consistent with the dominance of the sphinstanton --
anti-sphinstanton pair.

\section{Conclusions}

It often happens that classical solutions
determining the factorial
growth of the perturbation series have
physical significance (describe
tunneling) either in the original theory,
 or in the theory at unphysical
value of the coupling constant \cite{BBW,Lpt,BF,BPZJ,ZJ}.
Sphinstantons break this ''rule'' : while contributing
to large orders of perturbation theory, they do not have any other
apparent physical meaning. Another specific feature of these solutions
is that either single sphinstantons or sphinstanton -- anti-sphinstanton
pairs contribute to the asymptotics of perturbative expansions depending
on whether the number of negative modes around them is odd or even.
We have demonstrated this property by rather indirect analytical
calculations, as well as by the numerical study. The explicit mechanism
which makes the determinant about a single sphinstanton to vanish
in the case of odd number of negative  modes remains to be understood.

We are indebted to T. Banks, Yu. A. Kubyshin, D. T. Son  and P. G. Tinyakov
for helpful discussions.

\section*{Appendix}

In this Appendix we illustrate the technique of sect. 3 by a simpler
example of $O(n)$-symmetric system with the potential $V(r)$ shown in fig. 7,
$V(r) \simeq r^{2}/2$ if $r \rightarrow 0$.
The radial Hamiltonian reads
\[ H=-\frac{\hbar^{2}}{2}\frac{d^{2}}{dr^{2}}-\frac{n-1}{2r}\hbar^{2}
\frac{d}{dr}+V(r) \]
The relevant classical solution is bounce that starts at euclidean time
$t=-\infty$ at $r=0$ , reaches the turning point $r_{+}$ at $t=0$ and
returns to $r=0$. The action for the bounce is
\[ S_{B}=2 \int_{0}^{r_{+}}  dr \sqrt{2V(r)} \]
Repeating the arguments of sect. 3 we write the contribution of the  bounce
in the form analogous to eq. (\ref{21*}),
\[ (Tr \exp(-\beta H/\hbar))_{B}=
\left[ \int d\tau \sqrt{2V(R)} G(R,R;\tau) \right] \]
\begin{equation} \times
 \frac{2\beta \exp(-\beta n/2)}{\hbar^{n/2}\Gamma(n/2)}\exp(-S_{B}/\hbar)
\left(r_{+}\exp\left(\int_{0}^{r_{+}} dr[1/\sqrt{2V(r)}-
1/r]\right)\right)^{n} \label{34*}
\end{equation}
where $R$ is close to $r_{+}$. To evaluate Green's function $G(R,R;\tau)$
we point out that near $r_{+}$, the potential is well approximated by
the linear function. Retaining only leading terms in $\hbar$, we write the
Hamiltonian at $R<r<r_{+}$ in the following way,
\[ H=-\frac{\hbar^{2}}{2}\frac{d^{2}}{dy^{2}}+ay  \]
where $y=r_{+}-r$. The exact Green's function at coinciding arguments for
this Hamiltonian is
\[  G(y_{R},y_{R};\tau)=\frac{1}{\sqrt{2\pi \hbar\tau}} \exp(-S(y_{R},y_{R};
\tau)/\hbar) \]
where
\[ S(y_{R},y_{R};\tau)=a\tau y_{R} - \frac{a^{2}\tau^{3}}{24}-
\frac{4\sqrt{2a}y_{R}^{3/2}}{3} \]
\[ y_{R}=r_{+}-r \]
(We do not write the factor due to the ''quantum'' part of the Hamiltonian,
$H_{1}=-\frac{n-1}{2r}\hbar^{2}\frac{d}{dr}$, because this factor cancels out
precisely in the same way as in sect. 3.)
Performing the integration over $\tau$ by the saddle point technique, we
obtain at  small $y_{R}$
\begin{equation} \int d\tau \sqrt{2V(R)} G(R,R;\tau)=i
\label{35*} \end{equation}
According to eqs. (\ref{14+}), (\ref{14++}), the large order asymptotics
of the perturbation series for the ground state energy is found from
eqs. (\ref{34*}), (\ref{35*})
\begin{equation}
E^{(k)}=-\frac{k!k^{n/2-1}}{\pi\Gamma(n/2)S_{B}^{k+n/2}}
\left(r_{+}\exp\int_{0}^{r_{+}} dr[1/\sqrt{2V(r)}-1/r]\right)^{n} \label{35+}
\end{equation}
At $n=1$ this result coincides with the expression given in ref. \cite{ZJ}.
We know only one previous calculation at arbitrary $n$, namely, for
quartic potential \cite{BBW} $V(r)=r^{2}/2-r^{4}$. In that case we have
$r_{+}=1/\sqrt{2}$, $S_{B}=1/3$, and
\[ \int_{0}^{1/\sqrt{2}} dr [1/\sqrt{2V(r)}-1/r]=\ln 2 \]
So we find from  eq. (\ref{35+})
\[ E^{(k)}=-\frac{k!k^{n/2-1}}{\pi\Gamma(n/2)}3^{k+n/2}2^{n/2}\]
which coincides with the result of ref. \cite{BBW}. Thus, our technique
reproduces the known asymptotics.

\section*{Figure captions}

{\bf Fig. 1 : } Sphinstanton ( solid line ) at $n=2$ and its deformation
along the negative mode ( dashed line ) .
\newline
{\bf Fig. 2 : } Classical trajectory at finite but large $\beta$.
\newline
{\bf Fig. 3 : } Integration contour in complex $\lambda$ - plane.
\newline
{\bf Fig. 4 : } Coefficients $D_{k}$ defined by eq. (\ref{14A}) , as
functions of the order of perturbation theory, $k$ , at various
dimensionality of the sphere, $n$:
\newline
(a): $n=1.1$ to 1.9;
\newline
(b): $n=3.1$ to 3.9;
\newline
(c): $n=$ 2, 4, 6, 8, 10.
\newline
Dashed lines are asymptotic values,
 eq. (\ref{27*}).
\newline
{\bf Fig. 5 : } $D_{k}$ versus $k$ at integer odd dimensionality of the
sphere.
\newline
{\bf Fig. 6 : } The ratio (\ref{15B}) as function of the order of
 perturbation  theory $k$ at the dimensionalities of the sphere close
to $n=1$.The curves correspond to
\newline
1)  $\; n-1=10^{-3}$ ;
\newline
2)  $\; n-1=10^{-5}$ ;
\newline
3)  $\; n-1=10^{-7}$ ;
\newline
4)  $\; n-1=10^{-9}$ ;
\newline
5)  $\; n-1=10^{-11}$ ;
\newline
6)  $\; n-1=10^{-13}$ .

\newpage

\setlength{\unitlength}{0.240900pt}
\ifx\plotpoint\undefined\newsavebox{\plotpoint}\fi
\sbox{\plotpoint}{\rule[-0.200pt]{0.400pt}{0.400pt}}%
\begin{picture}(1600,1000)(0,0)
\font\gnuplot=cmr10 at 10pt
\gnuplot
\sbox{\plotpoint}{\rule[-0.200pt]{0.400pt}{0.400pt}}%
\put(220,776){$\theta_{r}$}
\put(220,835){$\pi$}
\put(220,230){$\theta_{s}$}
\put(220,108){0}
\put(270,900){$\theta$}
\put(260,776){\usebox{\plotpoint}}
\multiput(260.00,776.58)(2.593,0.499){119}{\rule{2.166pt}{0.120pt}}
\multiput(260.00,775.17)(310.505,61.000){2}{\rule{1.083pt}{0.400pt}}
\multiput(575.00,835.92)(2.593,-0.499){119}{\rule{2.166pt}{0.120pt}}
\multiput(575.00,836.17)(310.505,-61.000){2}{\rule{1.083pt}{0.400pt}}
\put(890,776){\usebox{\plotpoint}}
\multiput(890.00,774.92)(0.967,-0.492){19}{\rule{0.864pt}{0.118pt}}
\multiput(890.00,775.17)(19.207,-11.000){2}{\rule{0.432pt}{0.400pt}}
\multiput(911.58,762.85)(0.496,-0.522){39}{\rule{0.119pt}{0.519pt}}
\multiput(910.17,763.92)(21.000,-20.923){2}{\rule{0.400pt}{0.260pt}}
\multiput(932.58,739.74)(0.496,-0.862){39}{\rule{0.119pt}{0.786pt}}
\multiput(931.17,741.37)(21.000,-34.369){2}{\rule{0.400pt}{0.393pt}}
\multiput(953.58,702.79)(0.496,-1.152){39}{\rule{0.119pt}{1.014pt}}
\multiput(952.17,704.89)(21.000,-45.895){2}{\rule{0.400pt}{0.507pt}}
\multiput(974.58,653.84)(0.496,-1.443){39}{\rule{0.119pt}{1.243pt}}
\multiput(973.17,656.42)(21.000,-57.420){2}{\rule{0.400pt}{0.621pt}}
\multiput(995.58,592.81)(0.496,-1.758){39}{\rule{0.119pt}{1.490pt}}
\multiput(994.17,595.91)(21.000,-69.906){2}{\rule{0.400pt}{0.745pt}}
\multiput(1016.58,518.78)(0.496,-2.073){39}{\rule{0.119pt}{1.738pt}}
\multiput(1015.17,522.39)(21.000,-82.392){2}{\rule{0.400pt}{0.869pt}}
\multiput(1037.58,431.84)(0.496,-2.364){39}{\rule{0.119pt}{1.967pt}}
\multiput(1036.17,435.92)(21.000,-93.918){2}{\rule{0.400pt}{0.983pt}}
\multiput(1058.58,332.89)(0.496,-2.655){39}{\rule{0.119pt}{2.195pt}}
\multiput(1057.17,337.44)(21.000,-105.444){2}{\rule{0.400pt}{1.098pt}}
\multiput(1079.58,221.78)(0.496,-2.994){39}{\rule{0.119pt}{2.462pt}}
\multiput(1078.17,226.89)(21.000,-118.890){2}{\rule{0.400pt}{1.231pt}}
\multiput(1100.58,108.00)(0.496,2.994){39}{\rule{0.119pt}{2.462pt}}
\multiput(1099.17,108.00)(21.000,118.890){2}{\rule{0.400pt}{1.231pt}}
\multiput(1121.58,232.00)(0.496,2.655){39}{\rule{0.119pt}{2.195pt}}
\multiput(1120.17,232.00)(21.000,105.444){2}{\rule{0.400pt}{1.098pt}}
\multiput(1142.58,342.00)(0.496,2.364){39}{\rule{0.119pt}{1.967pt}}
\multiput(1141.17,342.00)(21.000,93.918){2}{\rule{0.400pt}{0.983pt}}
\multiput(1163.58,440.00)(0.496,2.073){39}{\rule{0.119pt}{1.738pt}}
\multiput(1162.17,440.00)(21.000,82.392){2}{\rule{0.400pt}{0.869pt}}
\multiput(1184.58,526.00)(0.496,1.758){39}{\rule{0.119pt}{1.490pt}}
\multiput(1183.17,526.00)(21.000,69.906){2}{\rule{0.400pt}{0.745pt}}
\multiput(1205.58,599.00)(0.496,1.443){39}{\rule{0.119pt}{1.243pt}}
\multiput(1204.17,599.00)(21.000,57.420){2}{\rule{0.400pt}{0.621pt}}
\multiput(1226.58,659.00)(0.496,1.152){39}{\rule{0.119pt}{1.014pt}}
\multiput(1225.17,659.00)(21.000,45.895){2}{\rule{0.400pt}{0.507pt}}
\multiput(1247.58,707.00)(0.496,0.862){39}{\rule{0.119pt}{0.786pt}}
\multiput(1246.17,707.00)(21.000,34.369){2}{\rule{0.400pt}{0.393pt}}
\multiput(1268.58,743.00)(0.496,0.522){39}{\rule{0.119pt}{0.519pt}}
\multiput(1267.17,743.00)(21.000,20.923){2}{\rule{0.400pt}{0.260pt}}
\multiput(1289.00,765.58)(0.967,0.492){19}{\rule{0.864pt}{0.118pt}}
\multiput(1289.00,764.17)(19.207,11.000){2}{\rule{0.432pt}{0.400pt}}
\multiput(1310.00,776.58)(2.663,0.496){37}{\rule{2.200pt}{0.119pt}}
\multiput(1310.00,775.17)(100.434,20.000){2}{\rule{1.100pt}{0.400pt}}

\put(1079,208){\usebox{\plotpoint}}
\put(1079.0,208.0){\rule[-0.200pt]{0.400pt}{5.059pt}}
\put(1079,168){\usebox{\plotpoint}}
\put(1079.0,168.0){\rule[-0.200pt]{0.400pt}{5.059pt}}
\put(1079,138){\usebox{\plotpoint}}
\put(1079.0,138.0){\rule[-0.200pt]{0.400pt}{5.059pt}}
\put(1079,108){\usebox{\plotpoint}}
\put(1079.0,108.0){\rule[-0.200pt]{0.400pt}{5.059pt}}

\put(1121,208){\usebox{\plotpoint}}
\put(1121.0,208.0){\rule[-0.200pt]{0.400pt}{5.059pt}}
\put(1121,168){\usebox{\plotpoint}}
\put(1121.0,168.0){\rule[-0.200pt]{0.400pt}{5.059pt}}
\put(1121,138){\usebox{\plotpoint}}
\put(1121.0,138.0){\rule[-0.200pt]{0.400pt}{5.059pt}}
\put(1121,108){\usebox{\plotpoint}}
\put(1121.0,108.0){\rule[-0.200pt]{0.400pt}{5.059pt}}

\put(1310,756){\usebox{\plotpoint}}
\put(1310.0,756.0){\rule[-0.200pt]{0.400pt}{5.059pt}}
\put(1310,716){\usebox{\plotpoint}}
\put(1310.0,716.0){\rule[-0.200pt]{0.400pt}{5.059pt}}
\put(1310,686){\usebox{\plotpoint}}
\put(1310.0,686.0){\rule[-0.200pt]{0.400pt}{5.059pt}}
\put(1310,656){\usebox{\plotpoint}}
\put(1310.0,656.0){\rule[-0.200pt]{0.400pt}{5.059pt}}
\put(1310,626){\usebox{\plotpoint}}
\put(1310.0,626.0){\rule[-0.200pt]{0.400pt}{5.059pt}}
\put(1310,596){\usebox{\plotpoint}}
\put(1310.0,596.0){\rule[-0.200pt]{0.400pt}{5.059pt}}
\put(1310,566){\usebox{\plotpoint}}
\put(1310.0,566.0){\rule[-0.200pt]{0.400pt}{5.059pt}}
\put(1310,536){\usebox{\plotpoint}}
\put(1310.0,536.0){\rule[-0.200pt]{0.400pt}{5.059pt}}
\put(1310,506){\usebox{\plotpoint}}
\put(1310.0,506.0){\rule[-0.200pt]{0.400pt}{5.059pt}}
\put(1310,476){\usebox{\plotpoint}}
\put(1310.0,476.0){\rule[-0.200pt]{0.400pt}{5.059pt}}
\put(1310,446){\usebox{\plotpoint}}
\put(1310.0,446.0){\rule[-0.200pt]{0.400pt}{5.059pt}}
\put(1310,416){\usebox{\plotpoint}}
\put(1310.0,416.0){\rule[-0.200pt]{0.400pt}{5.059pt}}
\put(1310,416){\usebox{\plotpoint}}
\put(1310.0,416.0){\rule[-0.200pt]{0.400pt}{5.059pt}}
\put(1310,386){\usebox{\plotpoint}}
\put(1310.0,386.0){\rule[-0.200pt]{0.400pt}{5.059pt}}
\put(1310,356){\usebox{\plotpoint}}
\put(1310.0,356.0){\rule[-0.200pt]{0.400pt}{5.059pt}}
\put(1310,326){\usebox{\plotpoint}}
\put(1310.0,326.0){\rule[-0.200pt]{0.400pt}{5.059pt}}
\put(1310,296){\usebox{\plotpoint}}
\put(1310.0,296.0){\rule[-0.200pt]{0.400pt}{5.059pt}}
\put(1310,266){\usebox{\plotpoint}}
\put(1310.0,266.0){\rule[-0.200pt]{0.400pt}{5.059pt}}
\put(1310,236){\usebox{\plotpoint}}
\put(1310.0,236.0){\rule[-0.200pt]{0.400pt}{5.059pt}}
\put(1310,206){\usebox{\plotpoint}}
\put(1310.0,206.0){\rule[-0.200pt]{0.400pt}{5.059pt}}
\put(1310,176){\usebox{\plotpoint}}
\put(1310.0,176.0){\rule[-0.200pt]{0.400pt}{5.059pt}}
\put(1310,146){\usebox{\plotpoint}}
\put(1310.0,146.0){\rule[-0.200pt]{0.400pt}{5.059pt}}
\put(1310,108){\usebox{\plotpoint}}
\put(1310.0,108.0){\rule[-0.200pt]{0.400pt}{5.059pt}}

\put(890,756){\usebox{\plotpoint}}
\put(890.0,756.0){\rule[-0.200pt]{0.400pt}{5.059pt}}
\put(890,716){\usebox{\plotpoint}}
\put(890.0,716.0){\rule[-0.200pt]{0.400pt}{5.059pt}}
\put(890,686){\usebox{\plotpoint}}
\put(890.0,686.0){\rule[-0.200pt]{0.400pt}{5.059pt}}
\put(890,656){\usebox{\plotpoint}}
\put(890.0,656.0){\rule[-0.200pt]{0.400pt}{5.059pt}}
\put(890,626){\usebox{\plotpoint}}
\put(890.0,626.0){\rule[-0.200pt]{0.400pt}{5.059pt}}
\put(890,596){\usebox{\plotpoint}}
\put(890.0,596.0){\rule[-0.200pt]{0.400pt}{5.059pt}}
\put(890,566){\usebox{\plotpoint}}
\put(890.0,566.0){\rule[-0.200pt]{0.400pt}{5.059pt}}
\put(890,536){\usebox{\plotpoint}}
\put(890.0,536.0){\rule[-0.200pt]{0.400pt}{5.059pt}}
\put(890,506){\usebox{\plotpoint}}
\put(890.0,506.0){\rule[-0.200pt]{0.400pt}{5.059pt}}
\put(890,476){\usebox{\plotpoint}}
\put(890.0,476.0){\rule[-0.200pt]{0.400pt}{5.059pt}}
\put(890,446){\usebox{\plotpoint}}
\put(890.0,446.0){\rule[-0.200pt]{0.400pt}{5.059pt}}
\put(890,416){\usebox{\plotpoint}}
\put(890.0,416.0){\rule[-0.200pt]{0.400pt}{5.059pt}}
\put(890,416){\usebox{\plotpoint}}
\put(890.0,416.0){\rule[-0.200pt]{0.400pt}{5.059pt}}
\put(890,386){\usebox{\plotpoint}}
\put(890.0,386.0){\rule[-0.200pt]{0.400pt}{5.059pt}}
\put(890,356){\usebox{\plotpoint}}
\put(890.0,356.0){\rule[-0.200pt]{0.400pt}{5.059pt}}
\put(890,326){\usebox{\plotpoint}}
\put(890.0,326.0){\rule[-0.200pt]{0.400pt}{5.059pt}}
\put(890,296){\usebox{\plotpoint}}
\put(890.0,296.0){\rule[-0.200pt]{0.400pt}{5.059pt}}
\put(890,266){\usebox{\plotpoint}}
\put(890.0,266.0){\rule[-0.200pt]{0.400pt}{5.059pt}}
\put(890,236){\usebox{\plotpoint}}
\put(890.0,236.0){\rule[-0.200pt]{0.400pt}{5.059pt}}
\put(890,206){\usebox{\plotpoint}}
\put(890.0,206.0){\rule[-0.200pt]{0.400pt}{5.059pt}}
\put(890,176){\usebox{\plotpoint}}
\put(890.0,176.0){\rule[-0.200pt]{0.400pt}{5.059pt}}
\put(890,146){\usebox{\plotpoint}}
\put(890.0,146.0){\rule[-0.200pt]{0.400pt}{5.059pt}}
\put(890,108){\usebox{\plotpoint}}
\put(890.0,108.0){\rule[-0.200pt]{0.400pt}{5.059pt}}

\put(850,123){$\tau_{1}$}
\put(969,78){$\tau_{1}+\tau_{2}$}
\put(1021,28){$\tau_{1}+\tau_{2}+\tau_{3}$}
\put(1320,123){$\beta$}
\put(1420,123){t}
\put(1121,58){\vector(0,1){50}}
\put(280,10){Fig.2.}

\put(260,776){\usebox{\plotpoint}}
\put(260.0,776.0){\rule[-0.200pt]{5.059pt}{0.400pt}}
\put(291,776){\usebox{\plotpoint}}
\put(291.0,776.0){\rule[-0.200pt]{5.059pt}{0.400pt}}
\put(333,776){\usebox{\plotpoint}}
\put(333.0,776.0){\rule[-0.200pt]{5.059pt}{0.400pt}}
\put(375,776){\usebox{\plotpoint}}
\put(375.0,776.0){\rule[-0.200pt]{5.059pt}{0.400pt}}
\put(417,776){\usebox{\plotpoint}}
\put(417.0,776.0){\rule[-0.200pt]{5.059pt}{0.400pt}}
\put(459,776){\usebox{\plotpoint}}
\put(459.0,776.0){\rule[-0.200pt]{5.059pt}{0.400pt}}
\put(501,776){\usebox{\plotpoint}}
\put(501.0,776.0){\rule[-0.200pt]{5.059pt}{0.400pt}}
\put(543,776){\usebox{\plotpoint}}
\put(543.0,776.0){\rule[-0.200pt]{5.059pt}{0.400pt}}
\put(585,776){\usebox{\plotpoint}}
\put(585.0,776.0){\rule[-0.200pt]{5.059pt}{0.400pt}}
\put(627,776){\usebox{\plotpoint}}
\put(627.0,776.0){\rule[-0.200pt]{5.059pt}{0.400pt}}
\put(669,776){\usebox{\plotpoint}}
\put(669.0,776.0){\rule[-0.200pt]{5.059pt}{0.400pt}}
\put(711,776){\usebox{\plotpoint}}
\put(711.0,776.0){\rule[-0.200pt]{5.059pt}{0.400pt}}
\put(753,776){\usebox{\plotpoint}}
\put(753.0,776.0){\rule[-0.200pt]{5.059pt}{0.400pt}}
\put(795,776){\usebox{\plotpoint}}
\put(795.0,776.0){\rule[-0.200pt]{5.059pt}{0.400pt}}
\put(837,776){\usebox{\plotpoint}}
\put(837.0,776.0){\rule[-0.200pt]{5.059pt}{0.400pt}}
\put(879,776){\usebox{\plotpoint}}
\put(879.0,776.0){\rule[-0.200pt]{5.059pt}{0.400pt}}
\put(921,776){\usebox{\plotpoint}}
\put(921.0,776.0){\rule[-0.200pt]{5.059pt}{0.400pt}}
\put(963,776){\usebox{\plotpoint}}
\put(963.0,776.0){\rule[-0.200pt]{5.059pt}{0.400pt}}
\put(1005,776){\usebox{\plotpoint}}
\put(1005.0,776.0){\rule[-0.200pt]{5.059pt}{0.400pt}}
\put(1047,776){\usebox{\plotpoint}}
\put(1047.0,776.0){\rule[-0.200pt]{5.059pt}{0.400pt}}
\put(1089,776){\usebox{\plotpoint}}
\put(1089.0,776.0){\rule[-0.200pt]{5.059pt}{0.400pt}}
\put(1131,776){\usebox{\plotpoint}}
\put(1131.0,776.0){\rule[-0.200pt]{5.059pt}{0.400pt}}
\put(1173,776){\usebox{\plotpoint}}
\put(1173.0,776.0){\rule[-0.200pt]{5.059pt}{0.400pt}}
\put(1215,776){\usebox{\plotpoint}}
\put(1215.0,776.0){\rule[-0.200pt]{5.059pt}{0.400pt}}
\put(1257,776){\usebox{\plotpoint}}
\put(1257.0,776.0){\rule[-0.200pt]{5.059pt}{0.400pt}}
\put(1299,776){\usebox{\plotpoint}}
\put(1299.0,776.0){\rule[-0.200pt]{5.059pt}{0.400pt}}
\put(1341,776){\usebox{\plotpoint}}
\put(1341.0,776.0){\rule[-0.200pt]{5.059pt}{0.400pt}}
\put(1383,776){\usebox{\plotpoint}}
\put(1383.0,776.0){\rule[-0.200pt]{5.059pt}{0.400pt}}
\put(260,232){\usebox{\plotpoint}}
\put(260.0,232.0){\rule[-0.200pt]{5.059pt}{0.400pt}}
\put(312,232){\usebox{\plotpoint}}
\put(312.0,232.0){\rule[-0.200pt]{5.059pt}{0.400pt}}
\put(354,232){\usebox{\plotpoint}}
\put(354.0,232.0){\rule[-0.200pt]{5.059pt}{0.400pt}}
\put(396,232){\usebox{\plotpoint}}
\put(396.0,232.0){\rule[-0.200pt]{5.059pt}{0.400pt}}
\put(438,232){\usebox{\plotpoint}}
\put(438.0,232.0){\rule[-0.200pt]{5.059pt}{0.400pt}}
\put(480,232){\usebox{\plotpoint}}
\put(480.0,232.0){\rule[-0.200pt]{5.059pt}{0.400pt}}
\put(522,232){\usebox{\plotpoint}}
\put(522.0,232.0){\rule[-0.200pt]{5.059pt}{0.400pt}}
\put(564,232){\usebox{\plotpoint}}
\put(564.0,232.0){\rule[-0.200pt]{5.059pt}{0.400pt}}
\put(606,232){\usebox{\plotpoint}}
\put(606.0,232.0){\rule[-0.200pt]{5.059pt}{0.400pt}}
\put(648,232){\usebox{\plotpoint}}
\put(648.0,232.0){\rule[-0.200pt]{5.059pt}{0.400pt}}
\put(690,232){\usebox{\plotpoint}}
\put(690.0,232.0){\rule[-0.200pt]{5.059pt}{0.400pt}}
\put(732,232){\usebox{\plotpoint}}
\put(732.0,232.0){\rule[-0.200pt]{5.059pt}{0.400pt}}
\put(774,232){\usebox{\plotpoint}}
\put(774.0,232.0){\rule[-0.200pt]{5.059pt}{0.400pt}}
\put(816,232){\usebox{\plotpoint}}
\put(816.0,232.0){\rule[-0.200pt]{5.059pt}{0.400pt}}
\put(858,232){\usebox{\plotpoint}}
\put(858.0,232.0){\rule[-0.200pt]{5.059pt}{0.400pt}}
\put(900,232){\usebox{\plotpoint}}
\put(900.0,232.0){\rule[-0.200pt]{5.059pt}{0.400pt}}
\put(942,232){\usebox{\plotpoint}}
\put(942.0,232.0){\rule[-0.200pt]{5.059pt}{0.400pt}}
\put(984,232){\usebox{\plotpoint}}
\put(984.0,232.0){\rule[-0.200pt]{5.059pt}{0.400pt}}
\put(1026,232){\usebox{\plotpoint}}
\put(1026.0,232.0){\rule[-0.200pt]{5.059pt}{0.400pt}}
\put(1068,232){\usebox{\plotpoint}}
\put(1068.0,232.0){\rule[-0.200pt]{5.059pt}{0.400pt}}
\put(1110,232){\usebox{\plotpoint}}
\put(1110.0,232.0){\rule[-0.200pt]{5.059pt}{0.400pt}}
\put(1152,232){\usebox{\plotpoint}}
\put(1152.0,232.0){\rule[-0.200pt]{5.059pt}{0.400pt}}
\put(1194,232){\usebox{\plotpoint}}
\put(1194.0,232.0){\rule[-0.200pt]{5.059pt}{0.400pt}}
\put(1236,232){\usebox{\plotpoint}}
\put(1236.0,232.0){\rule[-0.200pt]{5.059pt}{0.400pt}}
\put(1278,232){\usebox{\plotpoint}}
\put(1278.0,232.0){\rule[-0.200pt]{5.059pt}{0.400pt}}
\put(1320,232){\usebox{\plotpoint}}
\put(1320.0,232.0){\rule[-0.200pt]{5.059pt}{0.400pt}}
\put(1362,232){\usebox{\plotpoint}}
\put(1362.0,232.0){\rule[-0.200pt]{5.059pt}{0.400pt}}
\put(260,837){\usebox{\plotpoint}}
\put(260.0,837.0){\rule[-0.200pt]{5.059pt}{0.400pt}}
\put(291,837){\usebox{\plotpoint}}
\put(291.0,837.0){\rule[-0.200pt]{5.059pt}{0.400pt}}
\put(333,837){\usebox{\plotpoint}}
\put(333.0,837.0){\rule[-0.200pt]{5.059pt}{0.400pt}}
\put(375,837){\usebox{\plotpoint}}
\put(375.0,837.0){\rule[-0.200pt]{5.059pt}{0.400pt}}
\put(417,837){\usebox{\plotpoint}}
\put(417.0,837.0){\rule[-0.200pt]{5.059pt}{0.400pt}}
\put(459,837){\usebox{\plotpoint}}
\put(459.0,837.0){\rule[-0.200pt]{5.059pt}{0.400pt}}
\put(501,837){\usebox{\plotpoint}}
\put(501.0,837.0){\rule[-0.200pt]{5.059pt}{0.400pt}}
\put(543,837){\usebox{\plotpoint}}
\put(543.0,837.0){\rule[-0.200pt]{5.059pt}{0.400pt}}
\put(585,837){\usebox{\plotpoint}}
\put(585.0,837.0){\rule[-0.200pt]{5.059pt}{0.400pt}}
\put(627,837){\usebox{\plotpoint}}
\put(627.0,837.0){\rule[-0.200pt]{5.059pt}{0.400pt}}
\put(669,837){\usebox{\plotpoint}}
\put(669.0,837.0){\rule[-0.200pt]{5.059pt}{0.400pt}}
\put(711,837){\usebox{\plotpoint}}
\put(711.0,837.0){\rule[-0.200pt]{5.059pt}{0.400pt}}
\put(753,837){\usebox{\plotpoint}}
\put(753.0,837.0){\rule[-0.200pt]{5.059pt}{0.400pt}}
\put(795,837){\usebox{\plotpoint}}
\put(795.0,837.0){\rule[-0.200pt]{5.059pt}{0.400pt}}
\put(837,837){\usebox{\plotpoint}}
\put(837.0,837.0){\rule[-0.200pt]{5.059pt}{0.400pt}}
\put(879,837){\usebox{\plotpoint}}
\put(879.0,837.0){\rule[-0.200pt]{5.059pt}{0.400pt}}
\put(921,837){\usebox{\plotpoint}}
\put(921.0,837.0){\rule[-0.200pt]{5.059pt}{0.400pt}}
\put(963,837){\usebox{\plotpoint}}
\put(963.0,837.0){\rule[-0.200pt]{5.059pt}{0.400pt}}
\put(1005,837){\usebox{\plotpoint}}
\put(1005.0,837.0){\rule[-0.200pt]{5.059pt}{0.400pt}}
\put(1047,837){\usebox{\plotpoint}}
\put(1047.0,837.0){\rule[-0.200pt]{5.059pt}{0.400pt}}
\put(1089,837){\usebox{\plotpoint}}
\put(1089.0,837.0){\rule[-0.200pt]{5.059pt}{0.400pt}}
\put(1131,837){\usebox{\plotpoint}}
\put(1131.0,837.0){\rule[-0.200pt]{5.059pt}{0.400pt}}
\put(1173,837){\usebox{\plotpoint}}
\put(1173.0,837.0){\rule[-0.200pt]{5.059pt}{0.400pt}}
\put(1215,837){\usebox{\plotpoint}}
\put(1215.0,837.0){\rule[-0.200pt]{5.059pt}{0.400pt}}
\put(1257,837){\usebox{\plotpoint}}
\put(1257.0,837.0){\rule[-0.200pt]{5.059pt}{0.400pt}}
\put(1299,837){\usebox{\plotpoint}}
\put(1299.0,837.0){\rule[-0.200pt]{5.059pt}{0.400pt}}
\put(1341,837){\usebox{\plotpoint}}
\put(1341.0,837.0){\rule[-0.200pt]{5.059pt}{0.400pt}}
\put(1383,837){\usebox{\plotpoint}}
\put(1383.0,837.0){\rule[-0.200pt]{5.059pt}{0.400pt}}
\put(260,108){\usebox{\plotpoint}}
\put(260.0,108.0){\rule[-0.200pt]{0.400pt}{185.252pt}}
\put(260,108){\usebox{\plotpoint}}
\put(260.0,108.0){\rule[-0.200pt]{278.240pt}{0.400pt}}
\end{picture}

\newpage

\begin{picture}(1000,600)(0,200)
\put(0,500){\vector(1,0){1000}}
\put(500,300){\vector(0,1){400}}
\put(525,650){Im$\lambda$}
\put(900,525){Re$\lambda$}
\put(250,200){Fig.3}
\put(0,480){\line(1,0){420}}
\put(420,480){\line(0,1){40}}
\put(420,520){\vector(-1,0){420}}
\put(250,650){\vector(1,-1){150}}
\put(250,650){$\lambda=-1$}
\end{picture}
\newpage

\setlength{\unitlength}{0.240900pt}
\ifx\plotpoint\undefined\newsavebox{\plotpoint}\fi
\sbox{\plotpoint}{\rule[-0.200pt]{0.400pt}{0.400pt}}%
\begin{picture}(1500,900)(0,0)
\font\gnuplot=cmr10 at 10pt
\gnuplot
\sbox{\plotpoint}{\rule[-0.200pt]{0.400pt}{0.400pt}}%
\put(176.0,68.0){\vector(0,1){800}}
\put(176.0,415.0){\vector(1,0){1300}}
\put(176,415){\usebox{\plotpoint}}
\multiput(176.00,415.61)(11.402,0.447){3}{\rule{7.033pt}{0.108pt}}
\multiput(176.00,414.17)(37.402,3.000){2}{\rule{3.517pt}{0.400pt}}
\multiput(228.00,418.59)(3.058,0.489){15}{\rule{2.456pt}{0.118pt}}
\multiput(228.00,417.17)(47.903,9.000){2}{\rule{1.228pt}{0.400pt}}
\multiput(281.00,427.58)(1.682,0.494){29}{\rule{1.425pt}{0.119pt}}
\multiput(281.00,426.17)(50.042,16.000){2}{\rule{0.713pt}{0.400pt}}
\multiput(334.00,443.58)(1.249,0.496){39}{\rule{1.090pt}{0.119pt}}
\multiput(334.00,442.17)(49.737,21.000){2}{\rule{0.545pt}{0.400pt}}
\multiput(386.00,464.58)(0.987,0.497){51}{\rule{0.885pt}{0.120pt}}
\multiput(386.00,463.17)(51.163,27.000){2}{\rule{0.443pt}{0.400pt}}
\multiput(439.00,491.58)(0.870,0.497){57}{\rule{0.793pt}{0.120pt}}
\multiput(439.00,490.17)(50.353,30.000){2}{\rule{0.397pt}{0.400pt}}
\multiput(491.00,521.58)(0.766,0.498){65}{\rule{0.712pt}{0.120pt}}
\multiput(491.00,520.17)(50.523,34.000){2}{\rule{0.356pt}{0.400pt}}
\multiput(543.00,555.58)(0.737,0.498){69}{\rule{0.689pt}{0.120pt}}
\multiput(543.00,554.17)(51.570,36.000){2}{\rule{0.344pt}{0.400pt}}
\multiput(596.00,591.58)(0.717,0.498){71}{\rule{0.673pt}{0.120pt}}
\multiput(596.00,590.17)(51.603,37.000){2}{\rule{0.336pt}{0.400pt}}
\multiput(649.00,628.58)(0.723,0.498){69}{\rule{0.678pt}{0.120pt}}
\multiput(649.00,627.17)(50.593,36.000){2}{\rule{0.339pt}{0.400pt}}
\multiput(701.00,664.58)(0.759,0.498){67}{\rule{0.706pt}{0.120pt}}
\multiput(701.00,663.17)(51.535,35.000){2}{\rule{0.353pt}{0.400pt}}
\multiput(754.00,699.58)(0.870,0.497){57}{\rule{0.793pt}{0.120pt}}
\multiput(754.00,698.17)(50.353,30.000){2}{\rule{0.397pt}{0.400pt}}
\multiput(806.00,729.58)(1.046,0.497){47}{\rule{0.932pt}{0.120pt}}
\multiput(806.00,728.17)(50.066,25.000){2}{\rule{0.466pt}{0.400pt}}
\multiput(858.00,754.58)(1.682,0.494){29}{\rule{1.425pt}{0.119pt}}
\multiput(858.00,753.17)(50.042,16.000){2}{\rule{0.713pt}{0.400pt}}
\multiput(911.00,770.59)(4.740,0.482){9}{\rule{3.633pt}{0.116pt}}
\multiput(911.00,769.17)(45.459,6.000){2}{\rule{1.817pt}{0.400pt}}
\multiput(964.00,774.93)(3.925,-0.485){11}{\rule{3.071pt}{0.117pt}}
\multiput(964.00,775.17)(45.625,-7.000){2}{\rule{1.536pt}{0.400pt}}
\multiput(1016.00,767.92)(1.215,-0.496){41}{\rule{1.064pt}{0.120pt}}
\multiput(1016.00,768.17)(50.792,-22.000){2}{\rule{0.532pt}{0.400pt}}
\multiput(1069.00,745.92)(0.634,-0.498){79}{\rule{0.607pt}{0.120pt}}
\multiput(1069.00,746.17)(50.739,-41.000){2}{\rule{0.304pt}{0.400pt}}
\multiput(1121.58,703.61)(0.498,-0.596){101}{\rule{0.120pt}{0.577pt}}
\multiput(1120.17,704.80)(52.000,-60.803){2}{\rule{0.400pt}{0.288pt}}
\multiput(1173.58,640.86)(0.498,-0.822){103}{\rule{0.120pt}{0.757pt}}
\multiput(1172.17,642.43)(53.000,-85.430){2}{\rule{0.400pt}{0.378pt}}
\multiput(1226.58,552.94)(0.498,-1.099){101}{\rule{0.120pt}{0.977pt}}
\multiput(1225.17,554.97)(52.000,-111.972){2}{\rule{0.400pt}{0.488pt}}
\multiput(1278.58,438.01)(0.498,-1.382){103}{\rule{0.120pt}{1.202pt}}
\multiput(1277.17,440.51)(53.000,-143.505){2}{\rule{0.400pt}{0.601pt}}
\multiput(1331.58,290.95)(0.498,-1.705){103}{\rule{0.120pt}{1.458pt}}
\multiput(1330.17,293.97)(53.000,-176.973){2}{\rule{0.400pt}{0.729pt}}
\put(1246,385){$r_{+}$}
\put(1430,425){$r$}
\put(186,840){V}
\put(186,385){0}
\put(400,10){Fig.7}
\end{picture}
\end{document}